\documentclass[12pt]{iopart}

\usepackage{amssymb}
\usepackage{graphicx}
\usepackage{color}
\usepackage[dvipsnames]{xcolor}
\usepackage{url}
\usepackage{hyperref}

\usepackage{color}

\newcommand{\exciting}{\texttt{exciting}}
\newcommand{\ie}{{\it i.e.}}
\newcommand{\kpt}{\textbf{k}-point}
\newcommand{\kpts}{\textbf{k}-points}
\newcommand{\kgrid}{\textbf{k}-grid}
\newcommand{\rgkmax}{\texttt{rgkmax}}
\newcommand{\xs}{\texttt{xs}}
\newcommand{\rttddft}{\texttt{rt\_tddft}}

\begin{document}
	\submitto{Electronic Structure}
	
	\title{All-electron full-potential implementation of real-time TDDFT in \exciting}          

	\author{Ronaldo Rodrigues Pela$^{1,2}$, Claudia Draxl$^{1,2}$}

	\address{$^1$ Physics Department and IRIS Adlershof, Humboldt-Universität zu Berlin, Zum Gro\ss en Windkanal 2, 12489 Berlin}
	\address{$^2$ European Theoretical Spectroscopy Facility (ETSF)}
	\ead{ronaldo@physik-hu.berlin.de}

	\begin{abstract}
	Linearized augmented planewaves combined with local-orbitals (LAPW+lo) are arguably the most precise basis set to represent Kohn-Sham states. When employed within real-time time-dependent density functional theory (RT-TDDFT), they promise ultimate precision achievable for exploring the evolution of electronic excitations. In this work, we present an implementation of RT-TDDFT in the full-potential LAPW+lo code \exciting{}. We benchmark our results against those obtained by linear-response TDDFT with \exciting{} and by RT-TDDFT calculations with the Octopus code, finding a satisfactory level of agreement. To illustrate possible applications of our implementation, we have chosen three examples: the dynamic behavior of excitations in MoS$_2$ induced by a laser pulse, the third harmonic generation in silicon, and a pump-probe experiment in diamond.
	\end{abstract}

	\maketitle

	\section{Introduction}

TDDFT is a powerful tool to study excitations in many-body systems \cite{Runge_1984}. It is formally an exact theory that guarantees the existence of a one-to-one mapping between the evolution of the many-body wavefunction and a much less complicated object, namely the time-dependent density \cite{Runge_1984,Botti_2007}. Compared to many-body perturbation theory based on Green functions, TDDFT is computationally less demanding, which allows for studying systems containing up to hundreds or even thousands of atoms \cite{Andrade_2012,Jornet-Somoza_2015,Maitra_2016,Kolesov_2016,Morzan_2014}. 
	
In practical problems, TDDFT is usually employed either in the linear-response (LR) regime, where the density is evaluated in the frequency domain as a first-order response to an external perturbation potential, or directly in the time domain, by evolving the Kohn-Sham (KS) wavefunctions \cite{Maitra_2016}. Both approaches have advantages and limitations. Here, we focus on RT-TDDFT that allows, among others, for assessing the nonlinear regime, for studying the dynamics of excitations in response to ultra-fast laser pulses as observed in pump-probe spectroscopy, and for observing the coupling of electronic excitations to the vibrations of the nuclei in real time  \cite{BENDE2015103,Kolesov_2016,Morzan_2014,Yabana_2012,Meng_2008,Sato_2015,Lian_2018,Ojanpera_2012,Yamada_2019,Sato_2018,Miyamoto_2015}.
	
On the one hand, a key point for the accuracy of TDDFT and, thus a most meaningful comparison with experiments, is the choice of the time-dependent exchange-correlation (TD-XC) functional. Therefore, the improvement of existing approximations has been in the focus of recent active research \cite{Maitra_2016,Maitra_2017,Refaely-Abramson_2015,Pemmaraju_2019,Imamura_2015,Rigamonti_2015}. On the other hand, for a given XC functional, it is desirable to achieve ultimate numerical precision within a calculation. This usually translates into a question about the quality of the basis employed to represent KS wavefunctions. In this sense, with its all-electron, full-potential LAPW+lo approach, \exciting{} has proven to be one of the most accurate \emph{ab initio} codes \cite{Lejaeghereaad_2016}, capable of reaching micro-Hartree precision \cite{Gulans_2018}. Furthermore, it is a user-friendly code with a growing community, and offers plenty of tutorials on its implementations \cite{exciting_webpage}. Nevertheless, up to now, TDDFT is available in \exciting{} only in its LR formulation \cite{Gulans_2014} while RT-TDDFT is still missing. 

In this paper, we fill this gap and present our implementation of RT-TDDFT in the \exciting{} code. A summary of the theory behind and the details of the implementation are given in Section \ref{sec-implementation}. In Section \ref{sec-benchmarck}, we provide benchmarks by comparing our results with those obtained by the Octopus code \cite{Tancogne-Dejean_2020}, while more examples can be found in the Appendices. In Section \ref{sec-features}, we demonstrate three interesting features of our implementation, namely (1) an analysis of the behavior of MoS$_2$ after excitation with a laser pulse, (2) a study of the third harmonic generation in silicon, and (3) a simulation of a pump-probe experiment in diamond. Finally, in Section \ref{sec-conclusion}, we provide our conclusions.

	\section{Theory and Implementation}\label{sec-implementation}
We start by considering a physical system with periodic boundary conditions subjected to an electric field $\mathbf{E}(t)$ with spatial variations on a scale much larger than the periodicity, and hence treated constant in space. The standard inclusion of $\mathbf{E}(t)$ in the KS Hamiltonian would be by addition of a dipole term $\mathbf{r}\cdot\mathbf{E}(t)$ that would break the desired periodicity. In this case, it is advantageous to employ the velocity-gauge \cite{Bertsch_2000,Yabana_2006,Yabana_2012,Pemmaraju_2018}:

	\begin{equation}\label{eq-Hamiltonian}
	\hat{H}(\mathbf{r},t) = \frac{1}{2}\left(-\mathrm{i}\nabla + \frac{1}{c}\mathbf{A}(t)\right)^2
	+
	v_{KS}(\mathbf{r},t),
	\end{equation}
where $\mathbf{A}(t)$ is the vector potential, given in this gauge by $\mathbf{A}(t)=-c\int_0^t \mathbf{E}(t')dt'$, $c$ is the speed of light, and $v_{KS}$ is the TD-KS potential, a sum of the TD ionic, Hartree and XC potentials. We assume here the adiabatic approximation for the TD-XC potential \cite{2006TDFT,2012FoTD}. A KS wavefunction $\psi_{j\mathbf{k}}(\mathbf{r},t)$ labeled with index $j$ and associated to a wavevector $\mathbf{k}$ evolves as
	\begin{equation}\label{eq-TDKSequation}
	\hat{H}(\mathbf{r},t)\psi_{j\mathbf{k}} (\mathbf{r},t) = 
	\mathrm{i}\frac{\partial }{\partial t} \psi_{j\mathbf{k}} (\mathbf{r},t).
	\end{equation}

In \exciting{,} each KS wavefunction is expanded in terms of the  LAPW+lo basis set with coefficients $C_{j\mathbf{k}}(t)$:
	\begin{equation}
	| \psi_{j\mathbf{k}}(t)\rangle
	=
	\sum_{\mathbf{G}} C_{j\mathbf{k}\mathbf{G}}(t)| \phi_{\mathbf{G}+\mathbf{k}}\rangle + \sum_\gamma C_{j\mathbf{k}\gamma}(t)| \phi_{\gamma}\rangle,
	\end{equation}
where $| \phi_{\mathbf{G}+\mathbf{k}}\rangle$ and $| \phi_{\gamma}\rangle$ represent the LAPW part of the basis and the local orbitals (lo), respectively; $\mathbf{G}$ is a reciprocal lattice vector.
With the definition of the basis, the integration of Eq. (\ref{eq-TDKSequation}) reduces to the problem of finding how $C_{j\mathbf{k}}(t)$ propagate in time. For this, many approaches are available \cite{Castro_2004, GomesPueyo_2018, Dewhurst_2016}. Apart from the classical Runge-Kutta method for differential equations, we have implemented the following propagators: (i) simple exponential, (ii) exponential at the midpoint, (iii) approximate enforced time-reversal symmetry, (iv) commutator-free Magnus expansion of 4th order, and (v) exponential using a basis of the Hamiltonian's eigenvectors. To illustrate how our implementation works, we choose the simple exponential propagator, while  \ref{sec-Propagators} details the other cases. 

The evolution of KS wavefunctions in terms of a propagator $\hat{U}(t+\Delta t,t)$ is:	
	\begin{equation}
	| \psi_{j\mathbf{k}}(t+\Delta t)\rangle
	=
	\hat{U}(t+\Delta t,t)| \psi_{j\mathbf{k}}(t)\rangle,
	\end{equation}
	where 
	\begin{equation}
	\hat{U}(t+\Delta t,t)  =
	\hat{\mathcal{T}}\left[
	\exp\left(
	-\mathrm{i}\int_t^{t+\Delta t}
	\mathrm{d}\tau \hat{H}(\tau)
	\right)
	\right],
	\end{equation}
$\hat{\mathcal{T}}$ being the time-ordering operator. For the simple exponential propagator, $\hat{U}(t+\Delta t,t)$ is regarded purely as $\exp[-\mathrm{i}\Delta t\hat{H}(t)]$, the approximation being better, the smaller the time step $\Delta t$ is. For this propagator, the following expression dictates the evolution of $C_{j\mathbf{k}}(t)$:
	\begin{equation}\label{eq-exponentialCoeff}
	C_{j\mathbf{k}}(t+\Delta t)
	=
	\mathrm{exp}[-\mathrm{i}\Delta t S_{\mathbf{k}}^{-1}
	H_{\mathbf{k}}(t)] \; C_{j\mathbf{k}}(t).
	\end{equation}
	The matrix exponential in Eq. (\ref{eq-exponentialCoeff}) is approximated by a Taylor expansion up to the order defined by the user (4 is the default). $H_{\mathbf{k}}$ and $S_{\mathbf{k}}$ are, respectively, the Hamiltonian and overlap matrices in the basis set. Since our basis consists of two parts, {\it i.e.}, LAPWs and lo's, $H_{\mathbf{k}}$ and $S_{\mathbf{k}}$ usually have a block structure, exemplified for the Hamiltonian, as follows:
	\begin{equation}
	\left[
	\begin{array}{c|c}
	\langle \phi_{\mathbf{k}+\mathbf{G}} | \hat{H}(t) | \phi_{\mathbf{k}+\mathbf{G}'}\rangle & \langle \phi_{\mathbf{k}+\mathbf{G}} | \hat{H}(t) | \phi_{\gamma'}\rangle  \\ \hline
	\langle \phi_{\gamma} | \hat{H}(t) | \phi_{\mathbf{k}+\mathbf{G}'}\rangle & \langle \phi_{\gamma} | \hat{H}(t) | \phi_{\gamma'}\rangle
	\end{array}
	\right].
	\end{equation}
	In \exciting{,} each block is calculated employing different strategies, as described in Ref. \cite{Gulans_2014}. If $\mu$ and $\nu$ denote generic indexes that can be associated to a LAPW or lo, then, following Eq. (\ref{eq-Hamiltonian}), an arbitrary element $[H_{\mathbf{k}}(t)]_{\mu\nu} = \langle \phi_{\mu} | H(t) | \phi_{\nu} \rangle$ can be written as:
	\begin{eqnarray}
	[H_{\mathbf{k}}(t)]_{\mu\nu} &=& \frac{1}{2} \langle \nabla \phi_{\mu} | \nabla \phi_{\nu} \rangle + \langle \phi_{\mu} | v_{KS}(t) | \phi_{\nu}\rangle + \nonumber \\ &+& \frac{\mathbf{A}^2(t)}{2c^2}[S_{\mathbf{k}}]_{\mu\nu} - \frac{\mathrm{i}}{c}\mathbf{A}(t)\cdot \langle \phi_{\mu} | \nabla| \phi_{\nu} \rangle.
	\end{eqnarray}
	The procedure to obtain $\langle \nabla \phi_{\mu} | \nabla \phi_{\nu} \rangle$, $\langle \phi_{\mu} | v_{KS}(t) | \phi_{\nu}\rangle$, and $[S_{\mathbf{k}}]_{\mu\nu}$ is detailed in Ref. \cite{Gulans_2014}, whereas the momentum matrix elements, $\langle\phi_{\mu} | -\mathrm{i}\nabla| \phi_{\nu} \rangle$, are calculated as described in Ref. \cite{Vorwerk_2019}.

	\subsection{Dielectric and optical-conductivity tensors}
	The dielectric and optical properties are quantities that can be measured by various experimental probes. They are also of main interest in LR-TDDFT calculations. To obtain them with RT-TDDFT, the behavior of the macroscopic current density, $\mathbf{J}(t)$, in response to an external field needs to be evaluated. For the case of local and semilocal KS functionals, $\mathbf{J}(t)$ can be obtained as
	\begin{equation}\label{eq-J}
	\mathbf{J}(t) = \frac{\mathrm{i}}{\Omega}\sum_{j\mathbf{k}}w_{\mathbf{k}}f_{j\mathbf{k}}
	\left\langle \psi_{j\mathbf{k}}(t) \big|\nabla \big|\psi_{j\mathbf{k}}(t)\right\rangle
	-\frac{N\mathbf{A}(t)}{c\Omega},
	\end{equation}
	where $N$ is the number of valence electrons in the unit cell with volume $\Omega$, $w_{\mathbf{k}}$ is the weight of the considered \kpt{,} and $f_{j\mathbf{k}}$ is the occupation number of the corresponding KS state. After Fourier transform, we can obtain the components of the optical conductivity $\sigma$ and the dielectric tensor $\varepsilon$ as
	\begin{equation}\label{eq-sigma-epsilon}
	\sigma_{\alpha\beta}(\omega) = \frac{J_\alpha(\omega)}{E_\beta(\omega)}, \quad
	\varepsilon_{\alpha\beta}(\omega) = \delta_ {\alpha\beta}+ \frac{4\pi \mathrm{i}\sigma_{\alpha\beta}(\omega)}{\omega},
	\end{equation}
	where the indexes $\alpha$, $\beta$ mean the cartesian directions $x$, $y$ or $z$. It is convenient to consider an impulsive electric field $E_0\delta(t)$ in a specific direction $\alpha$ such that $E_\alpha(\omega) = E_0$. This is known as the transverse geometry \cite{Yabana_2012}. To understand its physical meaning, we look at the interface between the studied material and the vacuum, where the electric field comes from. Transverse geometry means that the field direction is parallel to the interface. Conversely, in the longitudinal geometry the electric field is perpendicular to the interface \cite{Yabana_2012}. In this case, the displacement fields $\mathbf{D}(t)$ inside and outside the system are related to each other through the surface charge, as given by the boundary conditions for electromagnetic fields \cite{Griffiths}.  Following Refs. \cite{Yabana_2012}, \cite{Bertsch_2000}, and \cite{Yabana_2006}, we consider the external component of the vector potential in the longitudinal geometry as given by $\mathbf{A}_{ext}(t) = -cD_0\theta(t)\mathbf{e}_\alpha$ or, equivalently, by $\mathbf{D}(t) = -(1/c)d\mathbf{A}_{ext}/dt = D_0\delta(t) \mathbf{e}_\alpha$. The induced vector potential, $\mathbf{A}_{ind}(t)$, is obtained from the current density as
	\begin{equation}
	\frac{d^2\mathbf{A}_{ind}}{dt^2} = 4 \pi c \mathbf{J}(t),
	\end{equation}
	and the total vector potential $\mathbf{A}(t)$, appearing in the Hamiltonian, is calculated as the sum $\mathbf{A}_{ext}(t)+\mathbf{A}_{ind}(t)$.

	\subsection{Number of excited electrons}
	A quantity of interest is the number of excited electrons after the interaction with a laser pulse \cite{Temnov_2006,Sato_2015,Sato_2014,Sokolowski_2000,Yabana_2012,Zheng_2019,Li_2016,Otobe_2008}. In RT-TDDFT, the occupation number $f_{j\mathbf{k}}$ of a KS state is kept fixed to its initial value. As the wavefunctions evolve, they are no longer eigenstates of $\hat{H}(t)$. It is possible to describe the number of excitations by projecting $| \psi_{i\mathbf{k}}(t)\rangle$ onto the adiabatic ground state of $\hat{H}(t)$'s eigenfunctions \cite{Otobe_2008,Sato_2015,Sato_2014} or onto the reference ground state at $t=0$  \cite{Otobe_2008,Yabana_2012,Li_2016}. We opt here for the latter. Therefore, for a given \kpt{,} we define the number of electrons that have been excited to an unoccupied KS state, labeled  $j$, as
	\begin{equation}
	m^e_{j\mathbf{k}} (t)= \sum_{i}
	f_{i\mathbf{k}}
	| \langle \psi_{j\mathbf{k}}(0)
	| \psi_{i\mathbf{k}}(t)\rangle |^2.
	\end{equation}
	Similarly, the number of holes created in an occupied KS state $j'$ can be specified as
	\begin{equation}
	m^h_{j'\mathbf{k}} (t)= f_{j'\mathbf{k}} - \sum_{i}
	f_{i\mathbf{k}}
	| \langle \psi_{j'\mathbf{k}}(0)
	| \psi_{i\mathbf{k}}(t)\rangle |^2.
	\end{equation}
Thus, the total number of excited electrons in a unit cell can be obtained by considering all the unoccupied states 
\begin{equation}
	N_{exc}(t)=
	\sum_{j\mathbf{k}}^{j\, unocc}
	w_\mathbf{k}  m^e_{j\mathbf{k}} (t) = \sum_{j'\mathbf{k}}^{j'\, occ}
	w_\mathbf{k}  m^h_{j'\mathbf{k}} (t) .
	\end{equation}

	\subsection{Parallelization}
\begin{figure}[htb]
		\centering
		\includegraphics[scale=1]{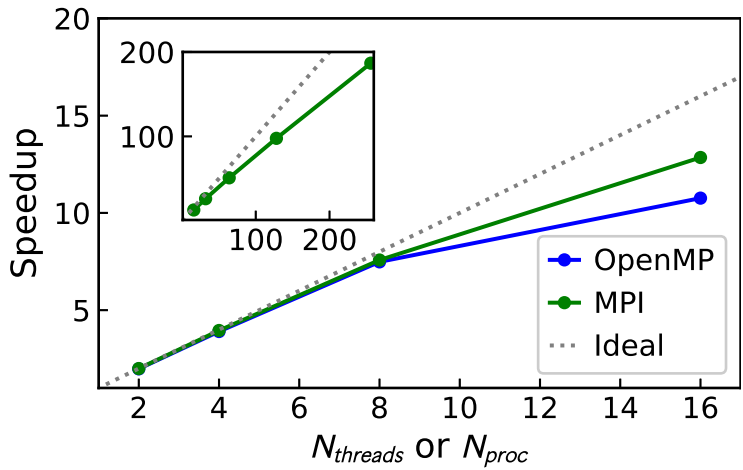}
		\caption{Comparison between the two parallelization schemes MPI and OpenMP for the current density in diamond using $16\times 16\times 16$ irreducible \kpts. The inset depicts the speedup by MPI when the job is distributed among several nodes. For comparison, the ideal scaling is indicated by the dashed line.}\label{fig-parallelization}
	\end{figure}

	We follow the same parallelization strategy as already adopted in other parts of \exciting, \ie{,} over \kpts{ }\cite{Gulans_2014}. In Fig. \ref{fig-parallelization}, we contrast the performance of two different levels of parallelization for calculations carried out on a single node with multiple processors: Open Multi-Processing (OpenMP) and Message Passing Interface (MPI). Although the speedup in both cases appears to be very close to the ideal one, MPI alone tends to be more efficient -- also when compared to a hybrid parallelization (using both OpenMP and MPI, not shown in the figure). In the inset of Fig. \ref{fig-parallelization}, we depict the speedup of MPI when the calculations are distributed over a higher number of nodes, still showing fairly close to ideal scaling (speedup of 187 for 256 processors).
	
	\subsection{Convergence behavior}

	In the following, we analyze the impact of the three most important parameters governing the precision of RT-TDDFT calculations, namely: the time step, the number of \kpts, and the size of the basis. For a given value $p$ of any of these parameters, we adopt the root-mean square error (RMSE)  
	\begin{equation}\label{eq-rms}
	\mathbb{E}_{p} = \sqrt{\frac{\int_0^T (j_p(t)-j_{ref}(t))^2\mathrm{d}t}{T}},
	\end{equation}
to address the convergence behavior. Here, $j_{ref}$ is a reference value for the current density, corresponding to the optimal parameter, and $T$ stands for the end-time, up to which the evolution of KS wavefunctions is considered.
	
\subsubsection{time step}
\begin{figure}[htb]
\centering
\includegraphics[scale=0.95]{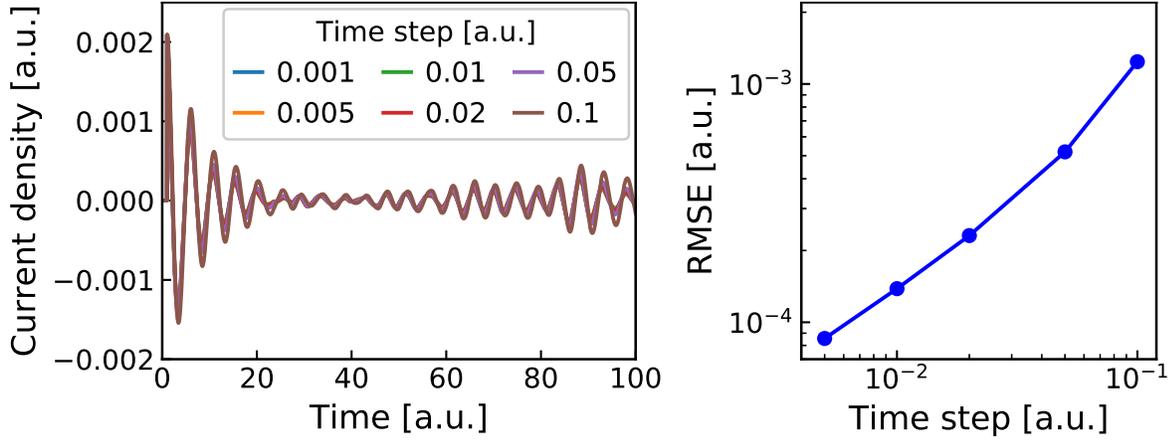}
\caption{Convergence behavior of the current density in diamond with respect to the chosen time step. The right panel shows the RMSE, taking the calculation with the smallest time step as reference.}\label{fig-convergence-C-tstep}
\end{figure}
	
We consider diamond under an impulsive displacement field along the [001], direction given by $D=0.02\delta(t-1)$ in atomic units (a.u.). In Fig. \ref{fig-convergence-C-tstep}, we show the  current-density response to this field in the same direction. On the one hand, the current density is apparently insensitive to the time step, as the various curves seem to coincide, suggesting swift convergence of the results with decreasing time-step. Interestingly, time steps above 0.2 a.u. lead to divergence. On the other hand, taking the current density obtained with a time step of 0.001 a.u. as $j_{ref}$ in Eq. (\ref{eq-rms}), the RMSE depicted on the right side of the figure shows that the calculations become indeed more precise when the time step is reduced. And there is actually no saturation behavior, \ie, the RMSE scales with the time step by a power law of 0.8. Although the value of the exponent depends on the material and on the method employed to propagate the wavefunctions, such power laws are found as a quite general trend, as already pointed out in Ref. \cite{Castro_2004}. A similar conclusion can be drawn in the case of silicon exposed to an electric field of two different forms (see Figs. \ref{fig-convergence-Si-tstep} and \ref{fig-convergence-Si-tstep-newlaser} in \ref{sec-Extra}).

\begin{figure}[bht]
\centering
\includegraphics[scale=0.95]{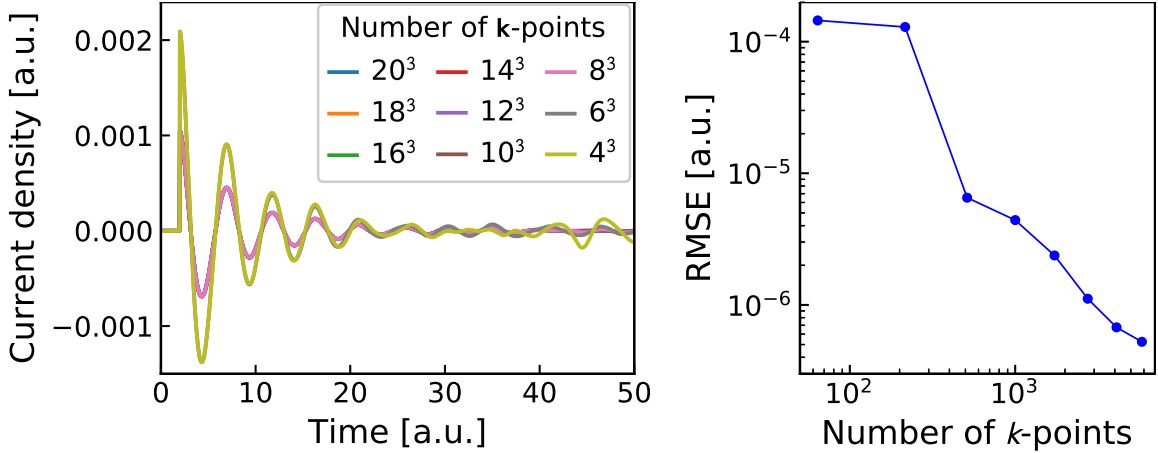}
\caption{Impact of the number of \kpts{} on the convergence of the current density. Results for diamond under the influence of an impulsive displacement field of $D=0.02\:\delta(t-2)$ applied along the [001] direction.}\label{fig-convergence-C-ngridk}
\end{figure}

\subsubsection{Number of \kpts}
	Analogous to ground-state calculations, the {\bf k}-grid has a direct impact on the quality of the current density (Eq. (\ref{eq-J})) and the time-dependent electronic density, calculated as
\begin{equation}
	n(\mathbf{r},t) = \sum_{j\mathbf{k}}w_{\mathbf{k}}f_{j\mathbf{k}}
	|\psi_{j\mathbf{k}}(\mathbf{r},t)|^2.
	\end{equation}
To illustrate its role, we consider diamond exposed to an impulsive displacement field $D=0.02\ \! \delta(t-2)$ a.u. along the [001] direction. Figure \ref{fig-convergence-C-ngridk} depicts how the number of \kpts{} affects the current density. Once more, the RMSE follows a power-law dependence on the investigated parameter (now, the number of \kpts), as seen in the right panel. The same applies to Figs. \ref{fig-convergence-Si-ngridk} and \ref{fig-convergence-Si-ngridk-newlaser}, where we consider Si excited by an impulsive field and a periodic function with a gaussian-like envelope, respectively.

\begin{figure}[thb]
\centering
\includegraphics[scale=0.935]{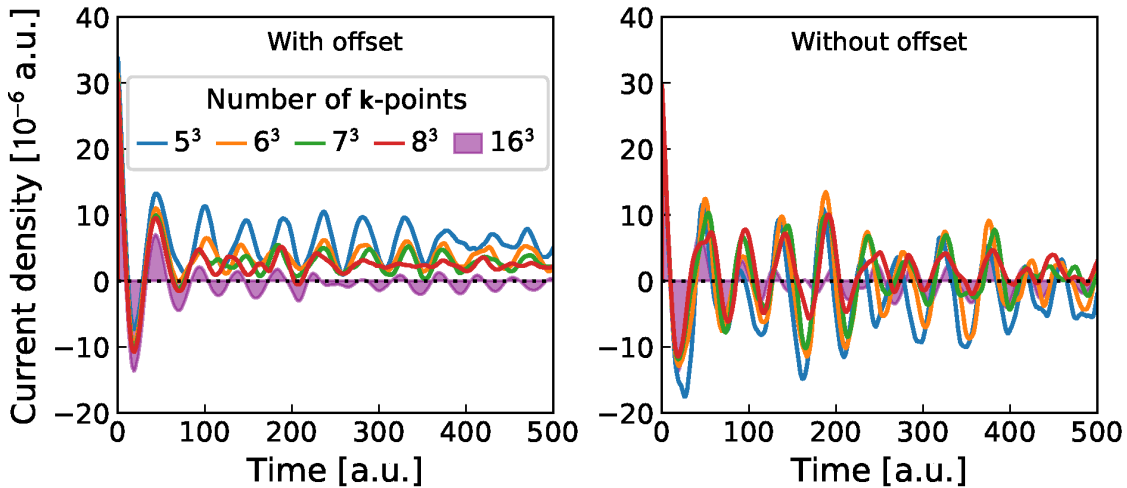}
\caption{Influence of the \kgrid{} on the current density with (left) and without (right) an offset.}\label{fig-convergence-Si-ngridk-off-curr}
\end{figure}
Another relevant aspect concerning the \kgrid{} is a possible offset that usually lowers the symmetry, leading to a set of symmetrically inequivalent \kpts{}. When the goal is, {\it e.g.}, to obtain the dielectric function, the offset helps to avoid symmetrically redundant contributions. To exemplify the effect of such offset, we take as a test case silicon exposed to an electric field along the [001] direction given by $E(t) = 0.001 \delta(t-0.16)$ a.u. Figure \ref{fig-convergence-Si-ngridk-off-curr} depicts on the left side the current density for different \kgrid{s} with an offset of $0.01 \ \!\mathbf{b}_1+0.45\ \!\mathbf{b}_2+0.37\ \!\mathbf{b}_3$ (where $\mathbf{b}_i$ are the reciprocal lattice vectors). The graph on the right side shows the case when no offset is taken into account. Comparing both graphs, we verify that, apart from a vertical shift, the current density converges faster with respect to the \kgrid{} when an offset is considered. The vertical shift signals that the offset induces an artificial long-time behavior $\displaystyle{ J(t\to\infty)}$ that does not converge to zero for coarser grids. This is not the case without an offset. Actually, in the linear regime, the summation in (\ref{eq-J}) should be ideally zero when the excitation field is removed. An offset may erroneously hamper cancellation of terms, this effect being much less pronounced when finer \kgrid{s} are considered.

\begin{figure}[bht]
\centering
\includegraphics[scale=.95]{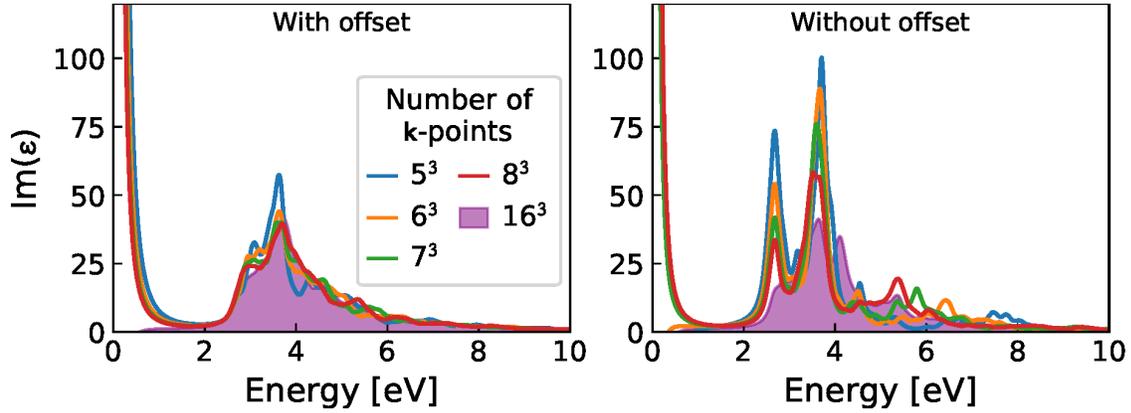}
\caption{Imaginary part of the dielectric function of Si varying the number of \kpts{} with (left) and without (right) an offset to break the symmetry.}\label{fig-convergence-Si-ngridk-off-eps}
\end{figure}
	Figure \ref{fig-convergence-Si-ngridk-off-eps} shows an equivalent comparison for the imaginary part of the dielectric function, calculated from the Fourier transform of the current density, as given in Eq. (\ref{eq-sigma-epsilon}). We note a ``fake'' plasmon at smaller frequencies (0-2~eV). This has already been reported in the literature as a consequence of the velocity-gauge \cite{Bertsch_2000,Pemmaraju_2018,Yabana_2012,Otobe_2009}. When no offset is included, the convergence with respect to the number of \kpts{} is slower. In contrast, when an offset is taken into account, calculations with a \kgrid{} of $8 \times 8 \times 8$ already show very similar results compared to doubling the points in each direction. Without the offset, transitions between valence and conduction band states tend to be sharper, and only a very high number of \kpts{} can describe those transitions that occur in the vicinity of \kpts{} with high-symmetry.

	\subsubsection{Basis}

\begin{figure}[htb]
		\centering
		\includegraphics[scale=.95]{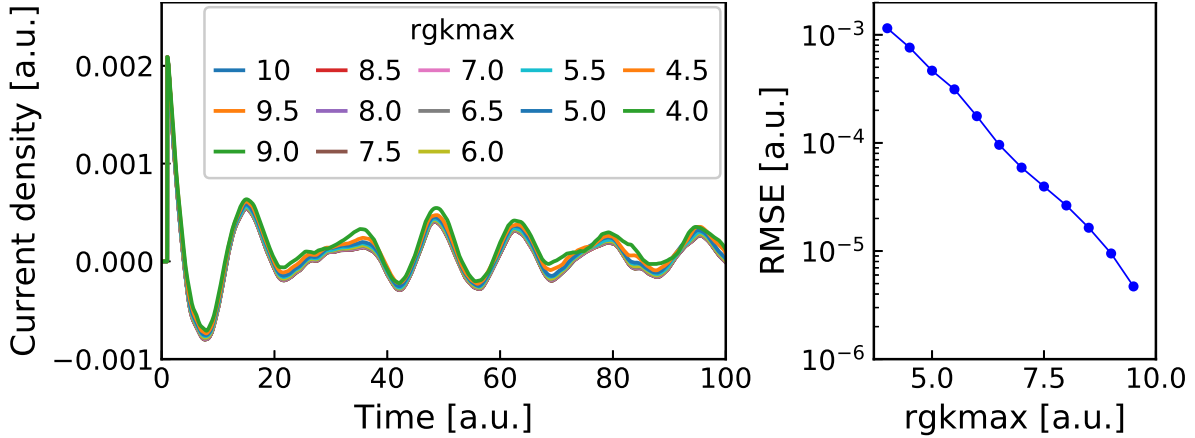}
		\caption{Convergence behavior of the current density in diamond for different values of \rgkmax, determining the basis-set size. The corresponding RMSE is displayed on the right.}\label{fig-convergence-C-rgkmax}
	\end{figure}

	In the case of LAPW+lo, the dimensionless parameter \rgkmax{} together with the number of lo's determine the quality of the basis. We need to inspect their impact on the convergence behavior separately. Starting with \rgkmax{}, we consider the current density in diamond exposed to an electric field $E(t) = 0.02 \delta(t-1)$ a.u. along [001]. From Fig. \ref{fig-convergence-C-rgkmax}, we conclude that the RMSE decreases exponentially when increasing \rgkmax. A similar behavior can also be observed in Fig. \ref{fig-convergence-Si-rgkmax-newlaser} (\ref{sec-Extra}) for silicon exposed to a sinusoidal electric field modulated by a gaussian-like envelope.

\begin{figure}[htb]
		\centering
		\includegraphics[scale=.95]{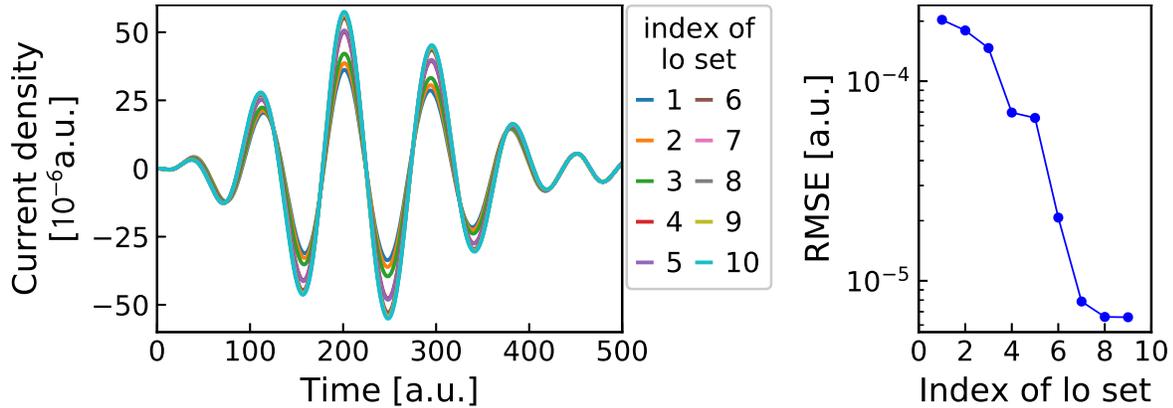}
		\caption{Convergence of the current density in silicon with increasing number of lo's (left). The following lo settings have been considered (number of lo's corresponding to angular momenta in parentheses) 1 (2 s, 1 p); 2 (2 s, 2 p) 3 (2 s, 2 p, 1 d); 4 (2 s, 2 p, 2 d); 5 (3 s, 2 p, 2 d); 6 (3 s, 3 p, 2 d); 7 (3 s, 3 p, 3 d); 8 (3 s, 4 p, 3 d); 9 (4 s, 4 p, 3 d); 10 (4 s, 4 p, 4 d). Right: corresponding RMSE.}\label{fig-convergence-Si-LO-newlaser}
	\end{figure}

To check the role of lo's, we consider silicon subjected to the electric field
	\begin{equation}\label{eq-GaussianLikeEnvelope}
	E(t) = E_m\sin^2\left[ \pi\frac{(t-t_0)}{ T_{pulse} } \right]\cos(\omega_0 t),
	\end{equation}
 along [001] for $t_0\le t \le T_{pulse}$, and 0 otherwise. This function describes a periodic wave with angular frequency $\omega_0$ modulated by a gaussian-like function, corresponding to a laser shape frequently employed in experiment. We choose, $E_m=1$, $\omega_0 = 0.0628$, $t_0=2$, and $T_{pulse}=452$, all quantities in a.u. In Fig. \ref{fig-convergence-Si-LO-newlaser}, we show how the current density changes when enhancing the basis with more lo's. Adding lo's with $p$ or $d$ character tends to improve the precision more than lo's with $s$ character. lo's with other character were found to have very little impact, thus these results are not shown here.  
	
	\section{Benchmark results}\label{sec-benchmarck}

	In this section, we present a benchmark of our implementation, contrasting the imaginary part of the dielectric function obtained with Eq. (\ref{eq-sigma-epsilon}) with that of the LR-TDDFT, employing the adiabatic local-density approximation (ALDA) as already implemented in \exciting{} \cite{Gulans_2014}. We also compare the current density obtained with our implementation with results from Octopus \cite{Tancogne-Dejean_2020}.

	\subsection{Comparison of RT- and LR-TDDFT: Dielectric function}
	
	\begin{figure}[htb]
		\centering
		\includegraphics[scale=1]{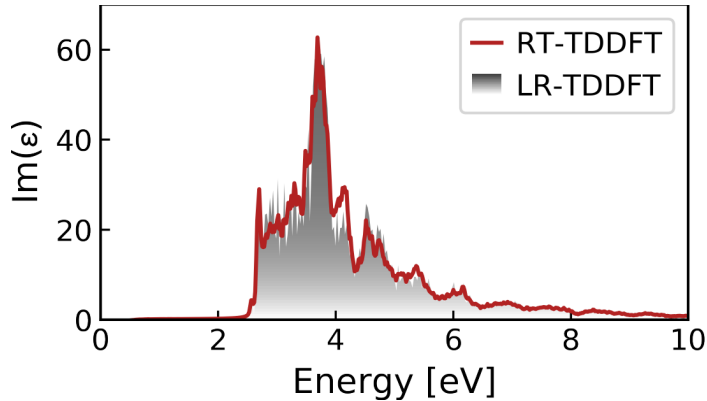}
		\caption{Imaginary part of the dielectric function of bulk silicon: Comparison between RT- and LR-TDDFT.}\label{fig-comparison-Si-RT-LR}
	\end{figure}
	
\begin{figure}[htb]
		\centering
	  	\includegraphics[scale=1]{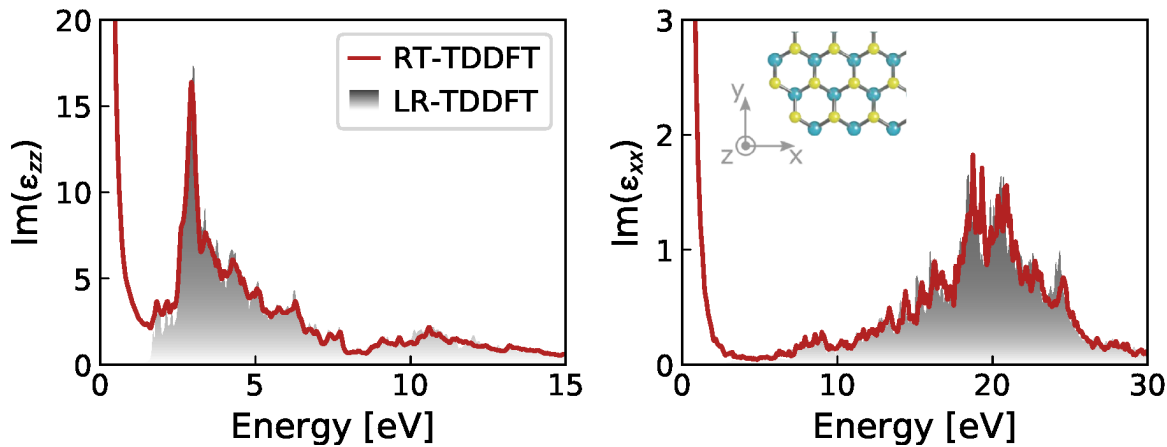}
	  	\caption{Out-of-plane ($zz$) (left) and in-plane ($xx$) (right) tensor-components of the imaginary part of the dielectric function of 2-dimensional MoS$_2$ obtained by RT-TDDFT in comparison with LR-TDDFT results}.\label{fig-comparison-MoS2-RT-LR}
	\end{figure}

	As prototypical materials for our initial benchmark, we choose silicon and 2-dimensional MoS$_2$. The imaginary part of their dielectric functions are given in Figs. \ref{fig-comparison-Si-RT-LR} and \ref{fig-comparison-MoS2-RT-LR}, respectively. In the RT-TDDFT calculations, we considered an impulsive electric field with an amplitude small enough to not induce deviations from the linear regime. Apart from the already commented ``fake'' plasmon in RT-TDDFT for smaller frequencies, we observe overall a remarkable agreement between both results. A similar comparison for diamond is provided in Fig. \ref{fig-comparison-C-RT-LR}.

	\subsection{Comparison with the Octopus code}

\begin{figure}[ht]
		\centering
		\includegraphics[scale=.95]{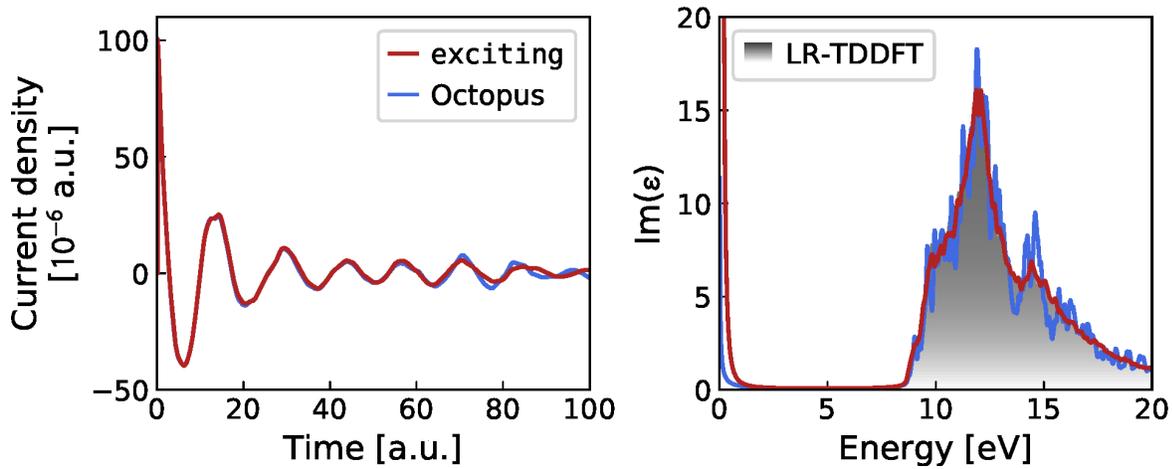}
		\caption{Comparison between the current density obtained with \exciting{} and Octopus for an impulsive electric field applied to cubic BN. The right panel compares corresponding results for the imaginary part of dielectric function to that obtained with the LR-TDDFT implementation of \exciting{} (gray shaded area).} \label{fig-comparison-octopus-BN}
	\end{figure}

	Octopus has been one of the first codes to evaluate the propagation of KS wavefunctions within the framework of RT-TDDFT \cite{Andrade_2012,Bertsch_2000,Castro_2004,Tancogne-Dejean_2020}. Hence, it is a most suitable package to benchmark our results, even though it employs a different scheme to solve the KS equations, namely pseudo-potentials combined with a real-space mesh \cite{Tancogne-Dejean_2020}.  On the left side of Fig. \ref{fig-comparison-octopus-BN}, we depict the current density in cubic BN as response to an impulsive electric field. The agreement between the results of \exciting{} and Octopus is impressive. On the right side, we provide the imaginary part of the dielectric function. This serves as well as a measure of how similar the Fourier-transforms of both curves are. We show also the result from our LR-TDDFT calculation, depicted as gray-shaded area. Once again, the agreement is excellent.
	
\begin{figure}[htb]
	\centering
	\includegraphics[scale=0.95]{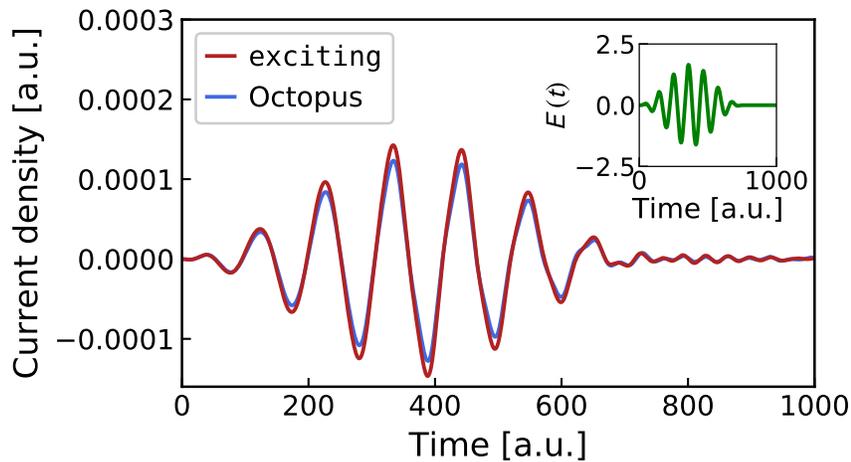}
	\caption{Comparison between the current density in Si, obtained with \exciting\ and Octopus. The inset depicts the applied electric field (in $10^{3}$~a.u.).}\label{fig-comparison-octopus-Si}
	\end{figure}

	 	 As a second benchmark, we depict in Fig. \ref{fig-comparison-octopus-Si} the current density in Si exposed to an electric field, whose expression follows Eq. (\ref{eq-GaussianLikeEnvelope}) and is shown in the inset, and compare the result to that obtained with Octopus. An interesting aspect here is that this field is strong enough to induce a nonlinear response, as it can be seen from the residual current density (after $t=700$ a.u., when the external field turns to zero). Also in this nonlinear regime, the agreement between \exciting{} and Octopus is very good. Similar agreement is found for SiC, see Fig. \ref{fig-comparison-octopus-SiC}.

\section{Implemented features}\label{sec-features}

The RT-TDDFT implementation naturally provides the evolution of the KS system, \ie, KS energies and wavefunctions, charge density, and total energy as functions of time. In this section, we choose three features of our implementation which highlight it as an interesting tool to aid the interpretation of experiments. 
	
\subsection{Excitation dynamics}
\begin{figure}[htb]
		\centering
		\includegraphics[scale=.94]{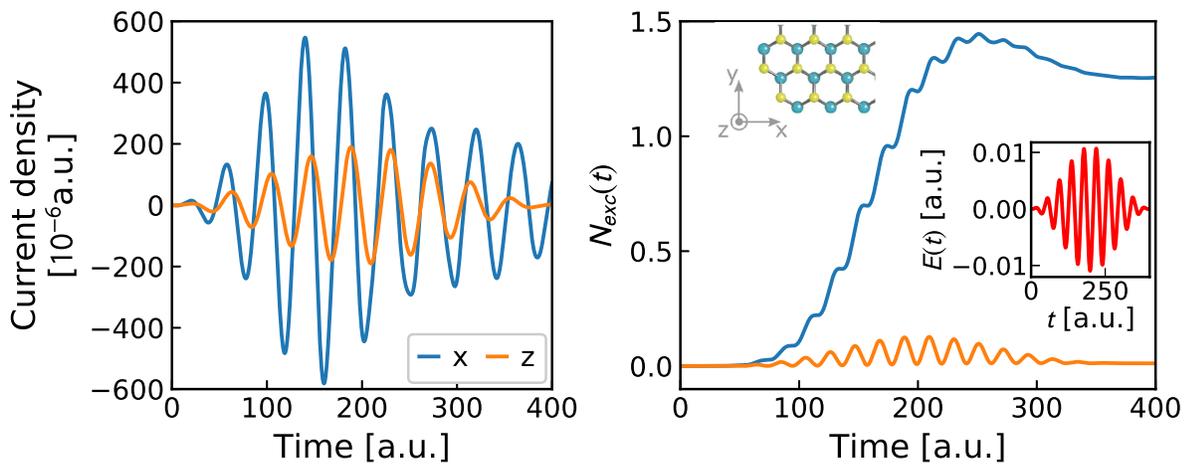}
		\caption{Left: Current density in MoS$_2$ under the action of electric fields parallel and perpendicular to the monolayer plane ($x$ and $z$ directions, respectively). Right: number of electrons per unit cell excited to the conduction band. The inset shows the time dependence of the applied field.}\label{fig-MoS2-CurrentDensity-Nexc}
	\end{figure}

	We start with the dynamics of an excitation in two-dimensional MoS$_2$ caused by a laser pulse with the electric field given by Eq. (\ref{eq-GaussianLikeEnvelope}) and plotted in the inset on the right panel of Fig. \ref{fig-MoS2-CurrentDensity-Nexc}. The pulse duration is set to $T_{pulse}=400$ a.u., the frequency to $\omega_0=0.15$~a.u. (corresponding to a photon energy of $4.08$~eV), and the peak intensity to $E_m=0.0107$~a.u. We carry out two calculations, one with the electric field parallel to the monolayer plane ($x$ direction), the other one perpendicular to it ($z$ direction). Figure \ref{fig-MoS2-CurrentDensity-Nexc} shows the current density in these two cases on the left, as well as the number of excited electrons on the right. 
	
	The current density is considerably higher in the case of in-plane polarization with a peak height being about 3 times larger. Some nonlinear effects are already observable. When we compare the number of excited electrons, we observe that, in the end, after the pulse is removed, 1.26 electrons per unit cell remain excited in the case of polarization along the $x$ direction, but two orders of magnitude less, \ie, 0.012, for the $z$ direction. We can understand this difference by the 2D nature of the material and can trace it back to the dielectric function (Fig. \ref{fig-comparison-MoS2-RT-LR}). The out-of-plane component $\varepsilon_{zz}$ at $4.08$~eV is much higher than the in-plane component which means that, at this frequency, MoS$_2$ can absorb electromagnetic waves with the electric field parallel to the monolayer plane much better than perpendicular to it.

\begin{figure}[htb]
		\centering
		\includegraphics[scale=1]{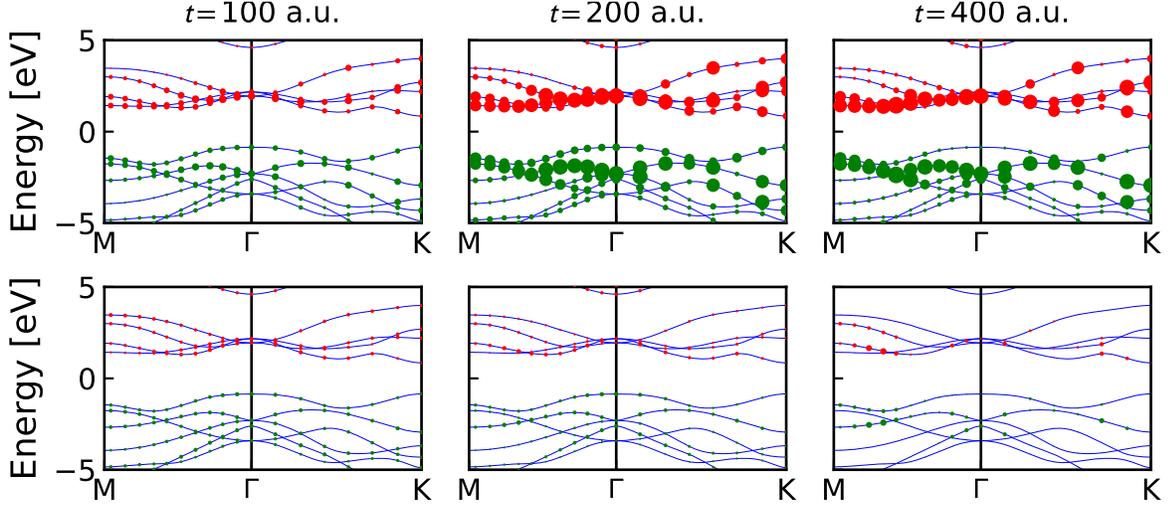}
		\caption{Band structure of MoS$_2$ along the $\Gamma$M and $\Gamma$K directions as the response to a laser-pulse with the electric field along (top panels) and perpendicular to (bottom panels) the MoS$_2$-plane. The circles in red (green) indicate the degree of population (depopulation) of the conduction (valence) bands at the specified times.}\label{fig-MoS2-Bands}
	\end{figure}

It is also interesting to observe from and to which bands the electrons are excited. In Fig. \ref{fig-MoS2-Bands} we show for three different times, \ie, $t=$ 100, 200 and 400~a.u. how the excitations are distributed over the {\bf k}-space. In the top panels, we provide the results for the $x$-polarization and in the bottom panels for the $z$-polarization.	
	In the case of in-plane polarization, some excitations are present at $t=$100~a.u.; many more electrons become excited at $t=$200~a.u.,  followed by a decrease thereafter. Interestingly, around the $\Gamma$ and $K$ points, the holes tend to be formed not on the valence-band top, but in deeper-lying bands, whereas at the $M$ point, the holes are predominantly at the top one. In the case of the perpendicular polarization, there is almost no difference between the excitations at times $t=$100~a.u. and $t=$200~a.u. In the end, only a few excitations remain, and they are not concentrated at the band edges, but rather in deeper- and higher-lying bands, respectively.

	\subsection{Non-linear response}
	We now analyze the response of silicon exposed to an electric field along the [001] direction whose expression follows Eq. (\ref{eq-GaussianLikeEnvelope}), with the parameters $\omega_0=0.0570$ a.u. (corresponding to a photon energy of 1.55 eV), $T_{pulse}=744$~a.u.~=~18.0~fs, and $t_0=0$, according to Ref. \cite{Yabana_2012}. The amplitude $E_m$ is varied so that we can observe a progression from the linear to the nonlinear regime.
	\begin{figure}[htb]
		\centering
		\includegraphics[scale=0.95]{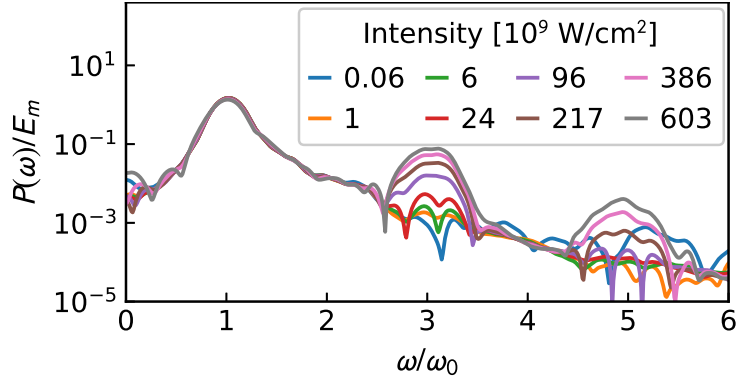}
		\caption{Fourier transform of the polarization field in silicon under the action of a laser pulse with fundamental frequency $\omega_0=0.0570$ a.u. (corresponding to a photon energy of 1.55 eV). The intensities shown in the legend refer to the values of the electromagnetic wave inside the bulk (which may differ from the nominal intensity applied to the sample -- see discussion in the text).}\label{fig-Si-3rdHarmonic}
	\end{figure}
	
	In Fig. \ref{fig-Si-3rdHarmonic}, we depict the Fourier-transform of the polarization field $P=(D-E)/(4\pi)$ normalized by the amplitude $E_m$ of the applied electric field. The intensity of the electromagnetic wave $I$ (in W/cm$^2$) is obtained from the amplitude as $I = 3.50941\times 10^{16}E_m^2$. It is important to recall that $E_m$ stands for the amplitude of the electric field inside the bulk material. Due to reflection at the interface and boundary conditions, the electric field generated by the exciting laser may be different from $E_m$, sometimes even two orders of magnitude higher \cite{Yabana_2012}. We can observe that, for intensities of $9.6\times 10^{10}$ W/cm$^{2}$ and higher, the third and even the fifth harmonic components are excited, and these components are obviously stronger the more intense the field is.

	\begin{figure}[htb]
		\centering
		\includegraphics[scale=0.95]{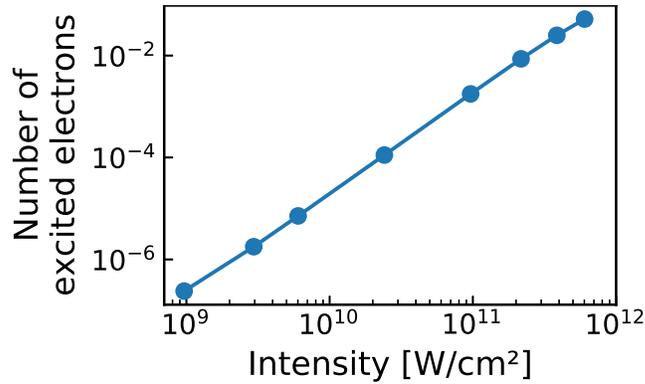}
		\caption{Number of excited electrons per unit cell in silicon as a function of the intensity of an external electric pulse of frequency $\omega_0=0.0570$ a.u. (corresponding to a photon energy of 1.55 eV). }\label{fig-Si-3rdHarmonic-NumberExcitedElectrons}
	\end{figure} 
	
We also evaluate the number of excited electrons per unit cell, $n_{ex}$, at a sufficient large time after the electric field has been switched off, as shown in Fig. \ref{fig-Si-3rdHarmonic-NumberExcitedElectrons}. We recognize that $n_{ex}$ is connected to the intensity $I$ of the electromagnetic wave by a power law, \ie,
	\begin{equation}
	n_{ex} = CI^n.
	\end{equation}
By means of a least square fit, we find $n=1.94$, which agrees with Ref. \cite{Otobe_2008}.

\subsection{Dielectric function after laser pulse}
\begin{figure}[htb]
		\centering
		\includegraphics[scale=1]{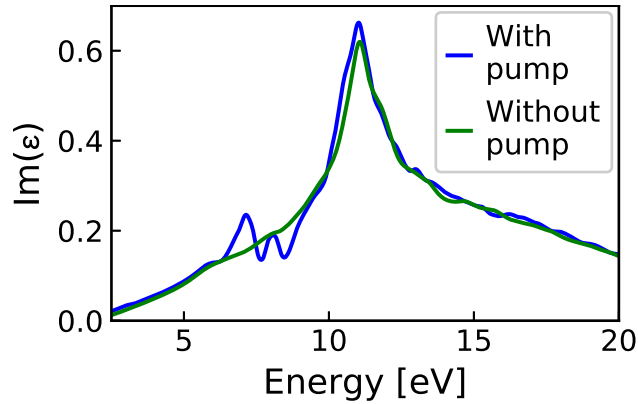}
		\caption{Imaginary part of the dielectric function of diamond probed after an electric field acting had been applied as pump (blue curve). For comparison, the curve expected without a pumping field is given in green.\label{fig-C-PumpProbe}}
	\end{figure}

We now simulate a pump-probe experiment, taking diamond as test material. The electric field of the pump pulse, given by Eq. (\ref{eq-GaussianLikeEnvelope}), has a gaussian-like envelope with width $T_{pulse}=644$~a.u., amplitude $E_m=2.2$~a.u., and fundamental frequency $\omega_0=0.1$~a.u (2.7~eV). At $t=700$~a.u., a weak impulsive electric field $E(t)=0.01 \delta(t-700)$ is applied as probe. We evaluate the dielectric function as indicated in Eq. (\ref{eq-sigma-epsilon}), but taking the current density as the difference between the values after the pump and the probe ($J_{pump-probe}$) and the pump pulse $J_{pump}$, as obtained from two separate calculations. In Fig. \ref{fig-C-PumpProbe}, we can identify that the main effect of the pump field is to change the absorption spectrum in the region between 6 and 11~eV,  especially around the third harmonic component (7.1~eV).

	\section{Conclusions}\label{sec-conclusion}

	In this paper, we have presented the implementation of RT-TDDFT in the full-potential LAPW+lo package \exciting{}, providing the underlying theory and details on the convergence behavior as well as parallelization performance. As benchmarks, we have compared our results with those obtained with the Octopus code as well as LR-TDDFT results from \exciting{}, finding excellent agreement in all cases. We have shown three examples of applications how our implementation could be used for the interpretation of experiments. These are the excitation dynamics of a material upon radiation with a laser pulse, the non-linear response of a material to laser pulses, and the dielectric function after a pump pulse. The implementation is included in the latest release, \exciting{} oxygen, and the code can be downloaded for free from the \exciting{} webpage \cite{exciting_webpage}. All data presented here are available in the NOMAD Repository \cite{Draxl_2019, nomad-doi} (DOI: 10.17172/NOMAD/2021.01.20-1).
	
	\ack 
	This work was supported by the Deutsche Forschungsgemeinschaft (DFG)- Projektnummer 182087777  - SFB 951. We thank Keith Gilmore, Santiago Rigamonti, Sven Lubeck and, Felix Henneke for the critical review of this manuscript. Alexander Buccheri and Sebastian Tillack are acknowledged for reviewing our code.

	\section*{References}
	\bibliographystyle{iopart-num}
	\bibliography{references}
	
	\appendix
	\section{Input}

	\begin{figure}[htb]
		\centering
		\includegraphics[scale=0.35]{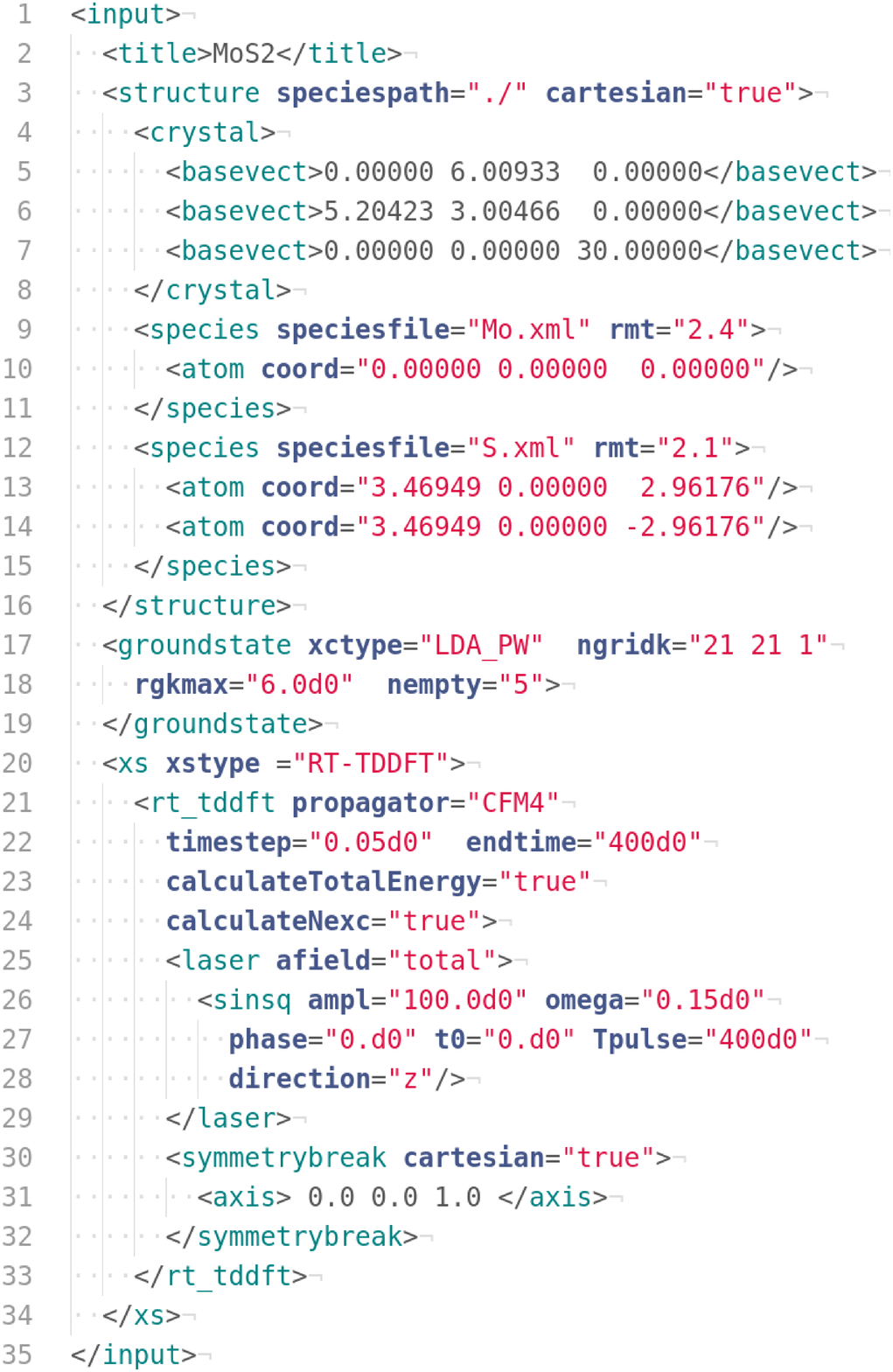}
		\caption{Example of input file ({\texttt{input.xml}}).}\label{fig-input-example}
	\end{figure}
	In Fig. \ref{fig-input-example}, we display as an example, the input file of MoS$_2$.For the RT-TDDFT calculations, the most important elements are captured by the element \rttddft\
	within the excited-state module \xs. Here, the input file defines as propagator CFM4 (commutator free Magnus of 4th order), with an evolution time starting at $0$ (default) up to $400$~a.u. ({\texttt{endtime}}), with steps of $0.05$~a.u. ({\texttt{timestep}}). The vector potential $\mathbf{A}(t)$, assuming the transverse geometry ({\texttt{afield="total"}}), is described by the element {\texttt{laser}}. In this case, we have a field applied along the $z$ axis, with gaussian-like envelope as in Eq. (\ref{eq-GaussianLikeEnvelope}), where $A_m=$100~a.u., $\omega_0=0.15$~a.u., $T_{pulse}=400$~a.u., $t_0=0$~a.u., and a null extra phase for the term $\cos(\omega_0 t)$. The element {\texttt{symmetry\_break}} defines an axis, given in Cartesian coordinates ({\texttt{cartesian="true"}}), to break the crystal symmetry. The full reference of input variables and their meaning is provided at the \exciting\ webpage \cite{exciting_webpage}.

	\section{Propagators}\label{sec-Propagators}
In this section, expressions for the propagator $\hat{U}(t+\Delta t,t)$ are provided. The derivations and assessments of their advantages or disadvantages can be found in Refs. \cite{Castro_2004}, \cite{GomesPueyo_2018}, and \cite{Dewhurst_2016}. We start with the most basic extension to the simple exponential propagator, namely the exponential at the midpoint:
	\begin{equation}
	\hat{U}(t+\Delta t,t)  =
	\exp\left[
	-\mathrm{i}\Delta t
	\hat{H}\left(t+\frac{\Delta t}{2}\right)
	\right],
	\end{equation}
	where, the extrapolation for obtaining the Hamiltonian $\hat{H}(t+f\Delta t)$ at future times is:
	\begin{equation}
	\hat{H}(t+f\Delta t)=
	(1+f)\hat{H}(t)
	-f\hat{H}(t-\Delta t).
	\end{equation}
	Another extension to the simple exponential propagator which keeps time-reversal symmetry to be approximately fulfilled is 
	\begin{equation}
	\hat{U}(t+\Delta t,t)  =
	\exp\left[-
	\mathrm{i}\frac{\Delta t}{2}
	\hat{H}\left(t+\Delta t\right)
	\right]
	\exp\left[
	-\mathrm{i}\frac{\Delta t}{2}
	\hat{H}\left(t\right)
	\right].
	\end{equation}
	Further improvement is provided by the so called commutator-Free Magnus expansion of 4th order:
	\begin{eqnarray}
    \hat{U}(t+\Delta t,t)  &=& 
	\exp\left[
	-\mathrm{i}\Delta t(
	\alpha_1\hat{H}(t_1)
	+
	\alpha_2\hat{H}(t_2))
	\right] \times \nonumber \\
	& \times &
	\exp\left[
	-\mathrm{i}\Delta t(
	\alpha_2\hat{H}(t_1)+
	\alpha_1\hat{H}(t_2))
	\right],
	\end{eqnarray}
	where $t_{1,2}=
	t+\left(
	\frac{1}{2}\mp
	\frac{\sqrt{3}}{6}
	\right) \Delta t$, and $\alpha_{1,2}=
	\frac{3\mp 2\sqrt{3}}{12}$.

A different approach is evaluating the exponential operator exactly rather than Taylor-expanding it. This can be done by taking into account an adiabatic basis formed by the eigenvectors of $\hat{H}(t)$. It means that for each $t$, we solve
	\begin{equation}\label{eq-adiabatic-basis}
	\hat{H}(t)| \phi_{j\mathbf{k}}(t)\rangle
	=
	\varepsilon_{j\mathbf{k}}(t)| \phi_{j\mathbf{k}}(t)\rangle	
	\end{equation}
and then expand
\begin{equation}
	| \psi_{j\mathbf{k}}(t)\rangle
	=
	\sum_i
	\alpha_{ij\mathbf{k}}(t)
	| \phi_{i\mathbf{k}}(t)\rangle,
	\label{eq:expansion}
	\end{equation}
where $\alpha_{ij\mathbf{k}}(t) = \langle \phi_{i}^{\mathbf{k}}(t)| \psi_{j}^{\mathbf{k}}(t)\rangle$. Since
\begin{equation}
	\hat{U}(t+\Delta t,t)
	|\phi_{m\mathbf{k}}(t)\rangle
	=
	\mathrm{e}^{-\mathrm{i}\varepsilon_{m\mathbf{k}}(t) \Delta t}
	|\phi_{m\mathbf{k}}(t)\rangle,
	\end{equation}
when considering $\hat{U}(t+\Delta t,t)$ in the form of the simple exponential propagator, the action of the propagator, using Eq. (\ref{eq:expansion}), is
    \begin{equation}
	\hat{U}(t+\Delta t,t)
	| \psi_{j\mathbf{k}}(t)\rangle
	=\sum_i \alpha_{ij\mathbf{k}}(t) \mathrm{e}^{-\mathrm{i}\varepsilon_{m\mathbf{k}}(t) \Delta t}
	|\phi_{m\mathbf{k}}(t)\rangle.
	\label{eq:action}
	\end{equation}
We further utilize then the expansion of $| \phi_{m\mathbf{k}}(t)\rangle$ in terms of our LAPW+lo basis.

Although the exponential operator is exactly obtained, \ie, without the need of a Taylor expansion, this approach now relies on an expansion in terms of the adiabatic basis and on the assumption of a simple exponential for the propagator. A first refinement can be provided if we employ the exponential at the midpoint, and then consider the adiabatic basis of $\hat{H}(t+\Delta t/2)$. 

Finally, it is also possible to employ the classical integrator of differential equations, the Runge-Kutta method, where we consider here the 4th order, \ie,
	\begin{equation}
	| \psi_{j\mathbf{k}}(t+\Delta t)\rangle
	= | \psi_{j\mathbf{k}}(t) \rangle
	- \frac{\mathrm{i}\Delta t}{6}(k_1+2k_2+2k_3+k_4)
	\end{equation}
	where
\begin{equation}
	k_1 = S_k^{-1}H_k(t)| \psi_{j\mathbf{k}}(t) \rangle,
	\end{equation}
	\begin{equation}
	k_2 = S_k^{-1}H_k\left(t + \frac{\Delta t}{2}\right) \left[| \psi_{j\mathbf{k}}(t) \rangle + k_1 \frac{\Delta t}{2}\right],
	\end{equation}
	\begin{equation}
	k_3 = S_k^{-1}H_k\left(t + \frac{\Delta t}{2}\right) \left[| \psi_{j\mathbf{k}}(t) \rangle + k_2 \frac{\Delta t}{2}\right],
	\end{equation}
	\begin{equation}
	k_4 = S_k^{-1}H_k(t+\Delta t)\left[| \psi_{j\mathbf{k}}(t) \rangle + k_3\Delta t \right].
	\end{equation}

	\section{Additional results}\label{sec-Extra}
	
	\subsection{Convergence behavior}
	\subsubsection{Time step}\label{subsubsec-timestep}

\begin{figure}[hbt]
		\includegraphics[scale=1.05]{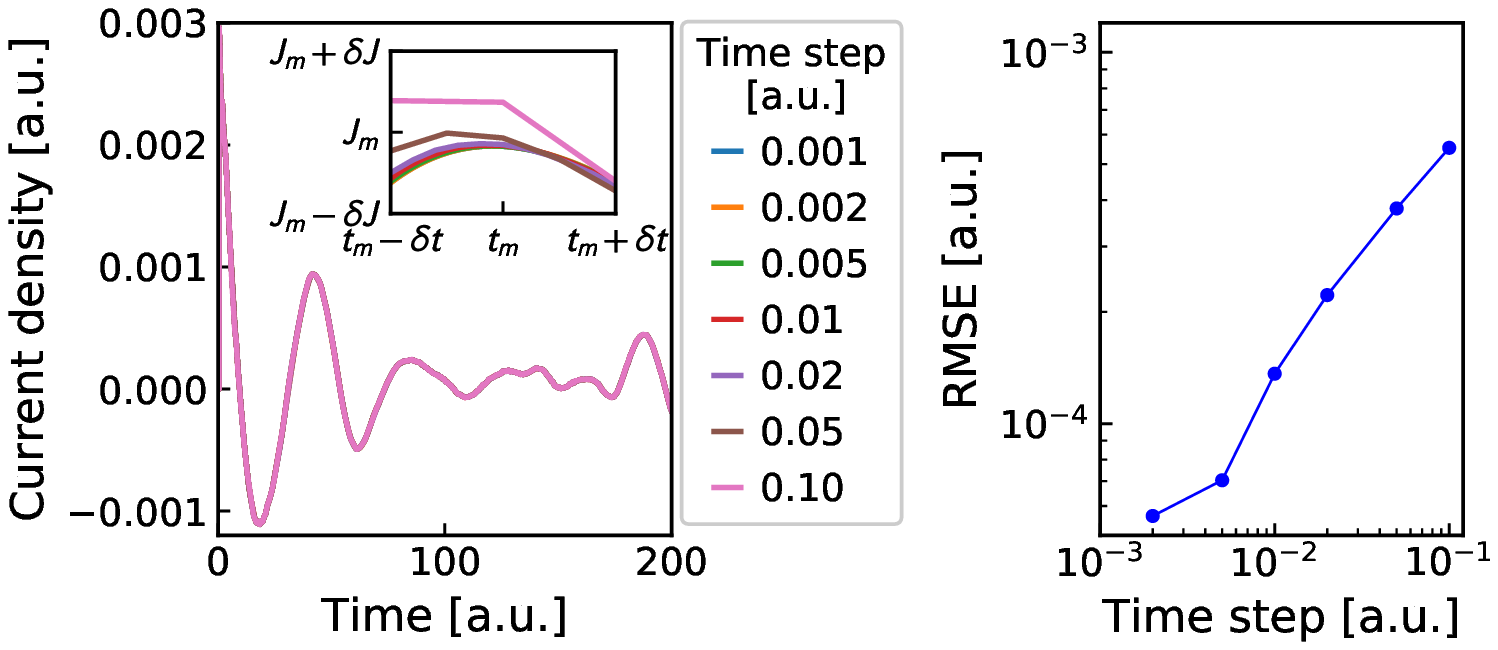}
		\caption{Convergence behavior of the current density in silicon with respect to time-step when an impulsive external field is applied (left). The inset amplifies the region around the first maximum $t_m$=41.5~a.u., with $J_m=9.43\times 10^{-4}$~a.u., $\delta J=1\times 10^{-6}$~a.u., and $\delta t=0.1$~a.u. The right panel displays the RMSE, taking the results with the smallest time-step as reference.}\label{fig-convergence-Si-tstep}
	\end{figure}

\begin{figure}[htb]
	    \centering
		\includegraphics[scale=1.1]{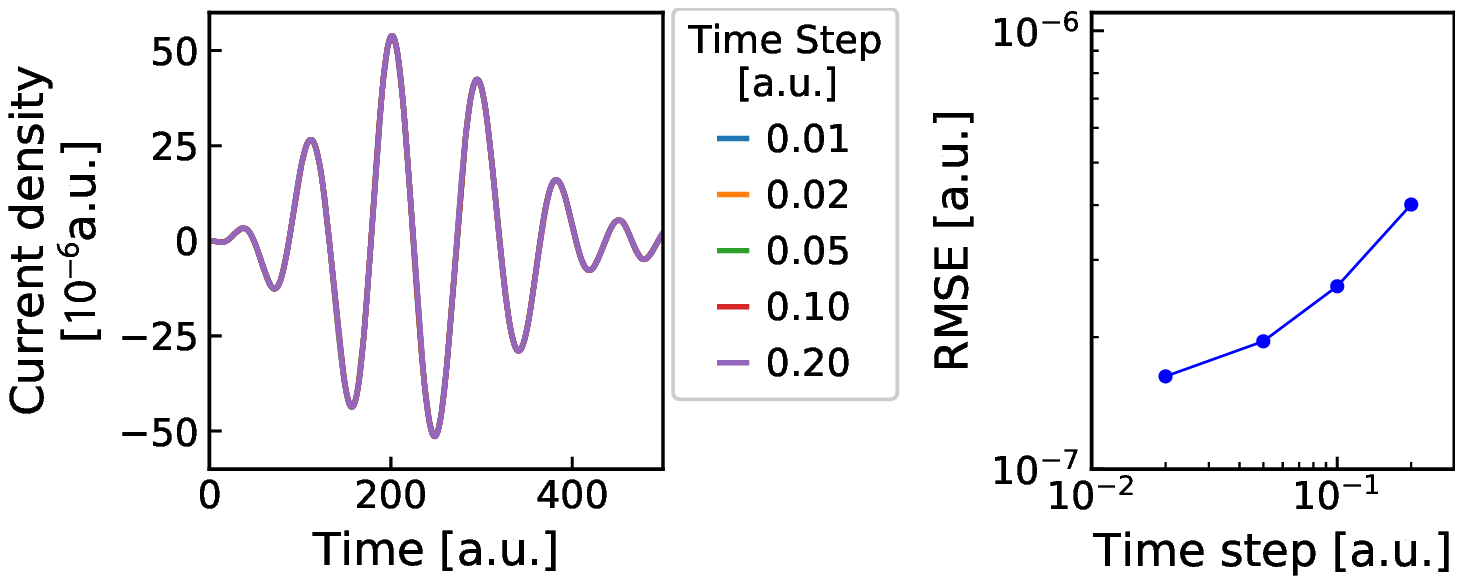}
		\caption{Convergence behavior of the current density in silicon with respect to time-step when an external field with gaussian-like envelope is applied. The RMSE is displayed on the right, taking the results obtained with a time-step $0.01$~a.u.as reference.}\label{fig-convergence-Si-tstep-newlaser}
	\end{figure}

	In Figs. \ref{fig-convergence-Si-tstep} and \ref{fig-convergence-Si-tstep-newlaser}, we show the convergence behavior of the current density in silicon, 
	probing the size of the time step after an external electric field is applied along the [001] direction. 
    In the first case, it is a delta function, $E(t) = 0.1\delta(t-0.5)$ a.u., while in the second case, it has a gaussian-like envelop, Eq. (\ref{eq-GaussianLikeEnvelope}), with the parameters $E_m=4.61\times 10^{-4}$, $\omega_0=0.0628$, $t_0=2.0$, and  $T_{pulse}=452$ (in atomic units).

	\subsubsection{\kpts}
\begin{figure}[htb]
	    \centering
		\includegraphics[scale=1.1]{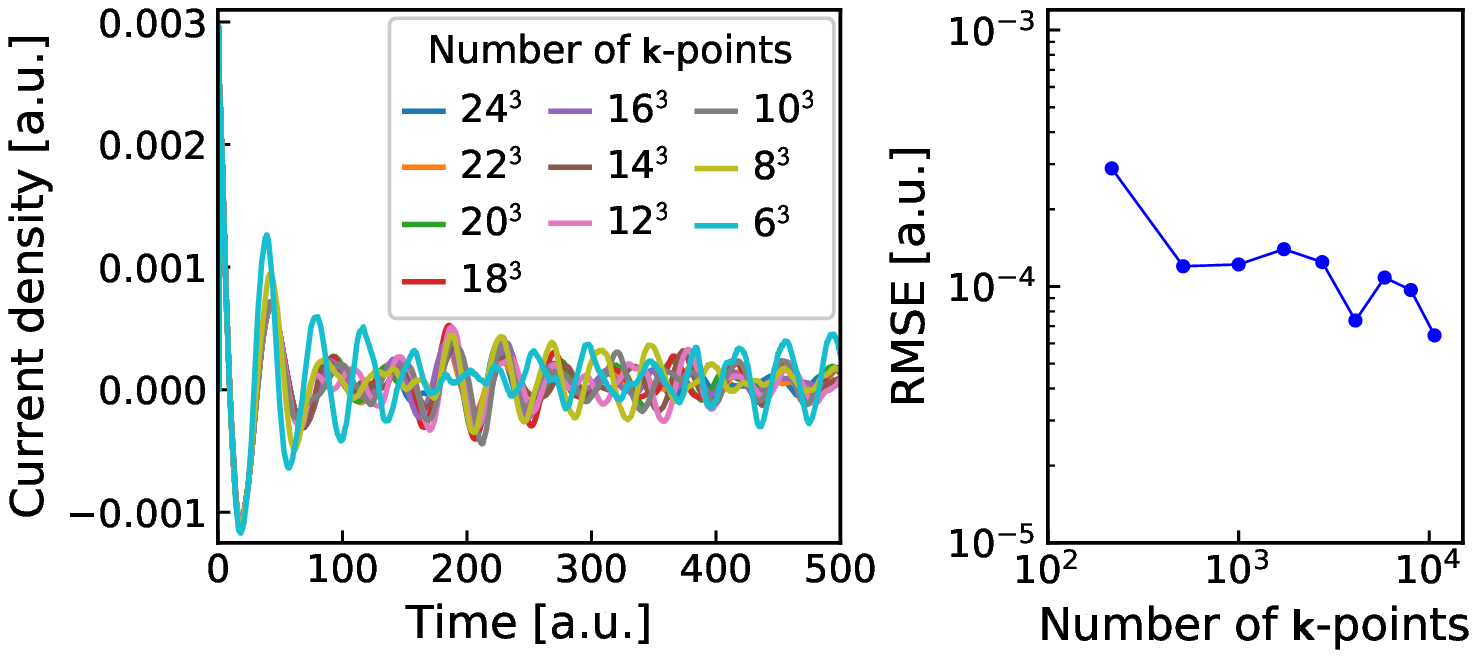}
		\caption{Impact of the number of \kpts{} on the current density in silicon, when an impulsive electric field is applied. The right panel shows the RMSE, taking the calculation with $24\times 24\times 24$ as reference. }\label{fig-convergence-Si-ngridk}
	\end{figure}

\begin{figure}[htb]
	    \centering
		\includegraphics[scale=1.05]{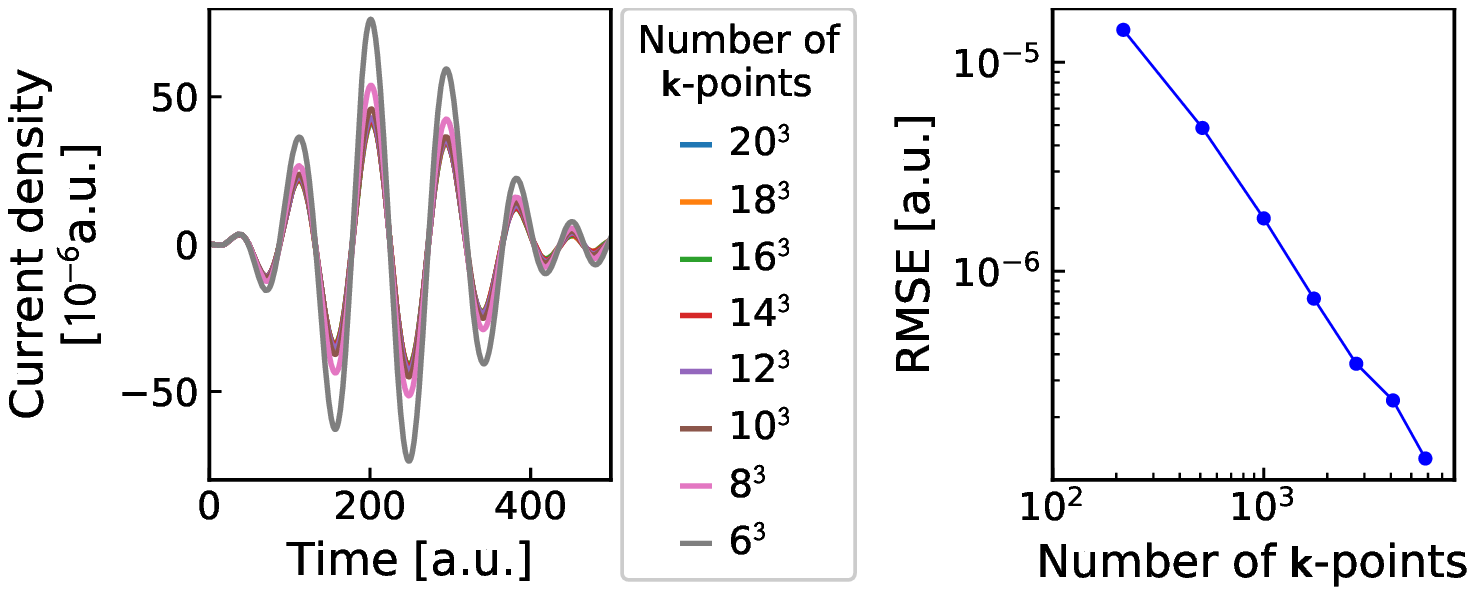}
		\caption{Convergence behavior of the current density in silicon with respect to the \kpts{},  where the external field has a gaussian-like envelope. The right panel shows the RMSE taking the calculation with highest a \kgrid{} of $20\times 20\times 20$ as reference. }\label{fig-convergence-Si-ngridk-newlaser}
	\end{figure}

	Figures \ref{fig-convergence-Si-ngridk} and \ref{fig-convergence-Si-ngridk-newlaser} display the convergence behavior of the current density in silicon with respect to the \kgrid{}. The external electric fields are the same as in \ref{subsubsec-timestep}.
	
\begin{figure}[htb]
		\centering
		\includegraphics[scale=1.05]{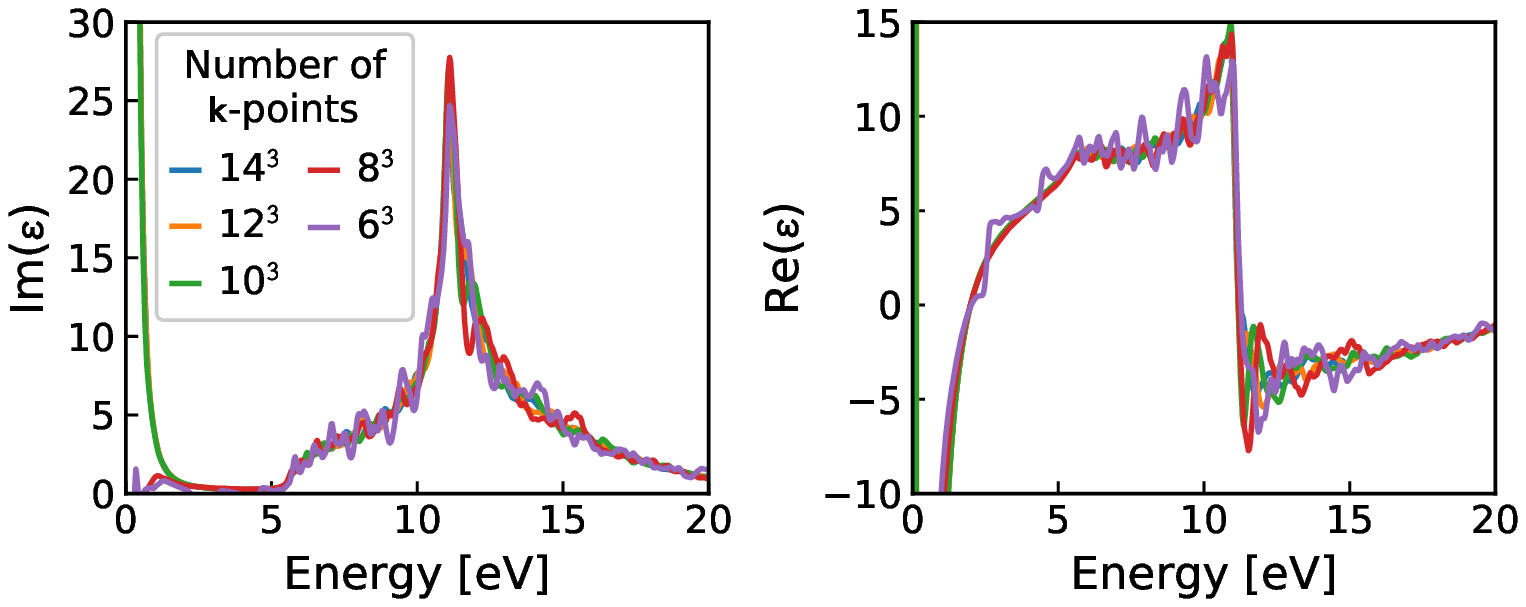}
		\caption{Impact of the number of \kpts{} on the convergence of the imaginary (left) and real (right) parts of the dielectric function of diamond.}\label{fig-convergence-C-ngridk-eps}
	\end{figure}

	Figure \ref{fig-convergence-C-ngridk-eps} shows the influence of the number of \kpts\ on the dielectric function of carbon. These results have been obtained from a Fourier transform of those given in Fig. \ref{fig-convergence-C-ngridk} (calculations up to a time of $5000$ a.u.).

	\subsubsection{Basis-set size}
\begin{figure}[hbt]
		\centering
		\includegraphics[scale=1]{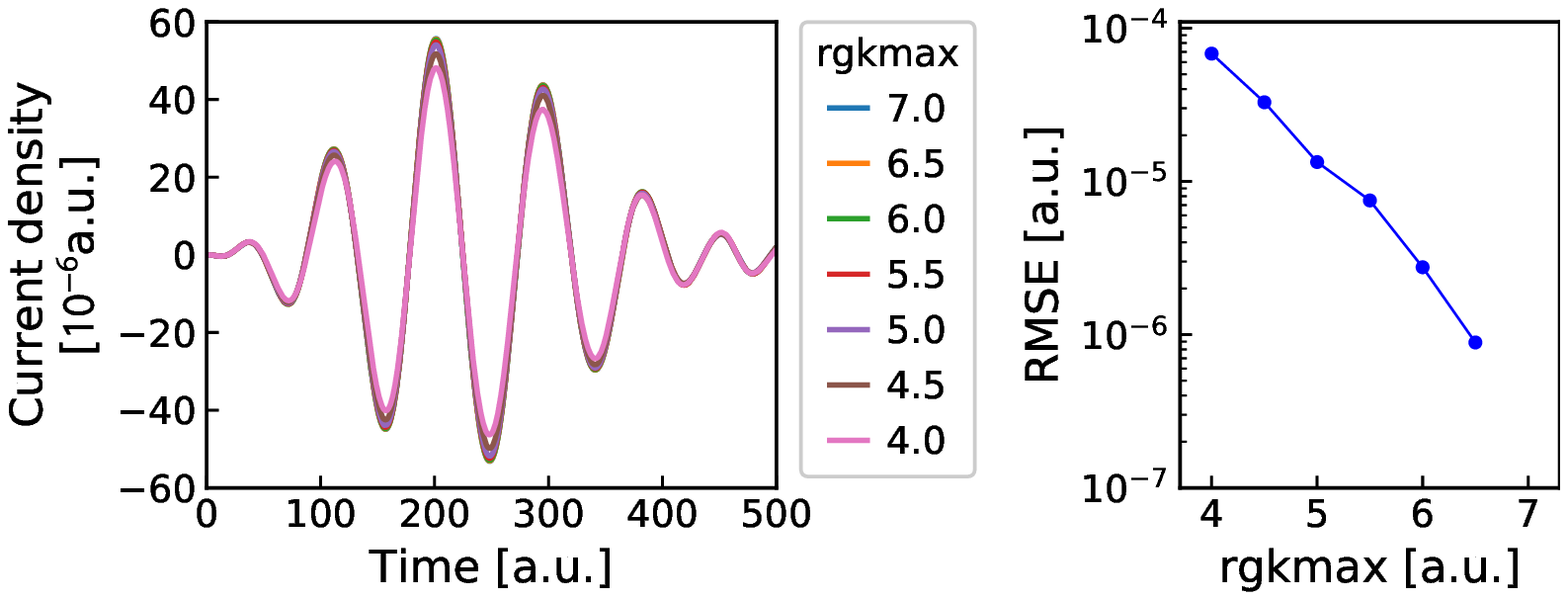}
		\caption{Convergence behavior of the current density in silicon with respect to the basis-set size, determined by the parameter \rgkmax{}. The external electric field has a gaussian-like envelop. The right panel depicts the RMSE, taking the calculation with highest \rgkmax\ as reference. }\label{fig-convergence-Si-rgkmax-newlaser}
	\end{figure}

	Figure \ref{fig-convergence-Si-rgkmax-newlaser} displays the impact of the choice of the parameter \rgkmax{} on the convergence behavior of the current density in silicon. The external electric field has a gaussian-like envelope, as given by Eq. (\ref{eq-GaussianLikeEnvelope}), with same parameters as in \ref{subsubsec-timestep}.

	\subsection{Benchmarks complementing Section \ref{sec-benchmarck}}
\begin{figure}[htb]
		\centering
		\includegraphics[scale=1]{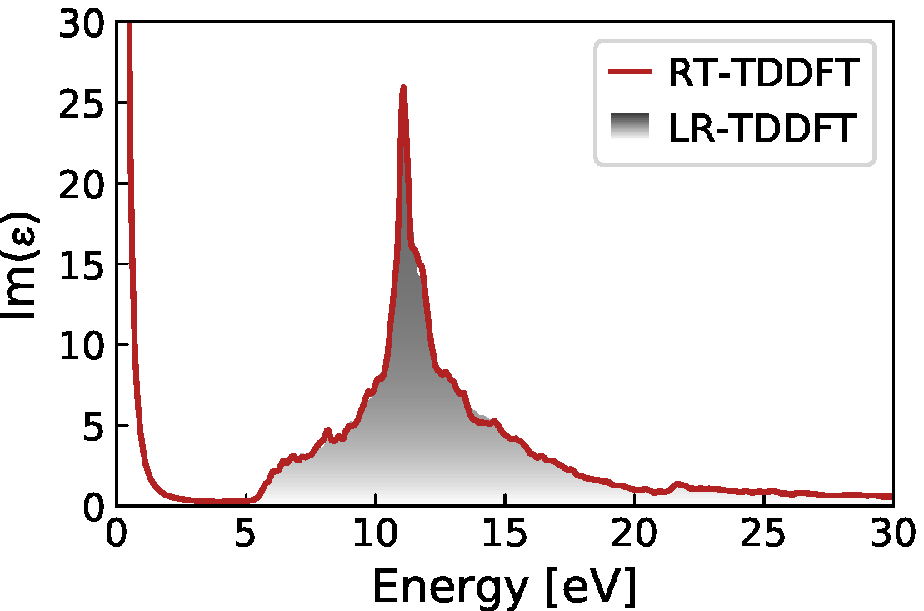}
		\caption{Imaginary part of the dielectric function of diamond: Comparison between RT- and LR-TDDFT.}\label{fig-comparison-C-RT-LR}
	\end{figure}

\begin{figure}[htb]
		\centering
		\includegraphics[scale=0.95]{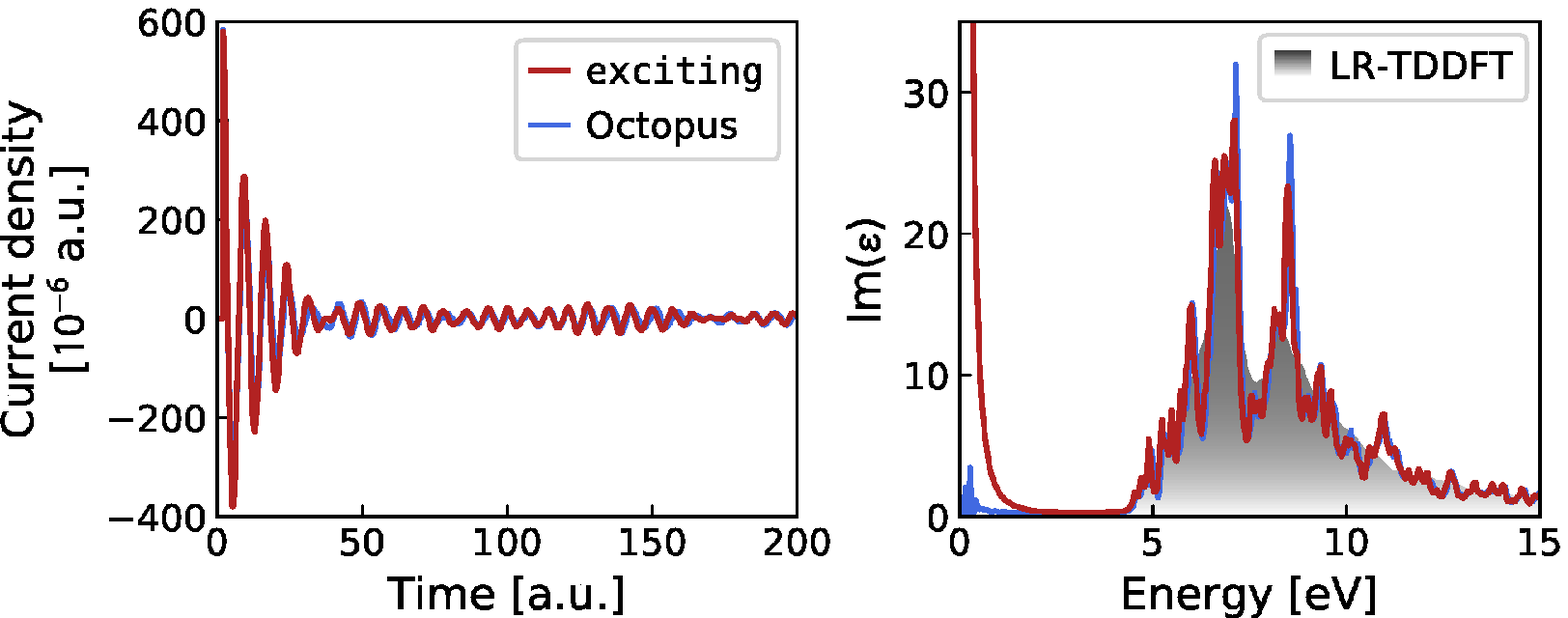}
		\caption{Current density in SiC, exposed to an impulsive field, as obtained with \exciting\ compared to the results of Octopus. The right panel compares the imaginary part of the dielectric function obtained from the RT-TDDFT implementations of both codes to the LR-TDDFT result obtained with \exciting.}\label{fig-comparison-octopus-SiC}
	\end{figure}

	Figure \ref{fig-comparison-C-RT-LR} displays how the imaginary part of the dielectric function of diamond obtained with RT-TDDFT compares with that from LR-TDDFT.
	In Fig. \ref{fig-comparison-octopus-SiC}, we compare the current density in SiC obtained with \exciting{} to the result by Octopus. The external field is a delta function applied along the [001] direction, i.e., $D(t) = 0.01\delta(t-2)$ [a.u.]. The right side of the figure shows how the imaginary part of the dielectric function obtained from the RT-TDDFT calculations of the two codes compares with LR-TDDFT obtained with \exciting.
	
\end{document}